\documentclass[namedreferences]{solarphysics}
\usepackage[hyperref,optionalrh,solaromanenum]{spr-sola-addons} 

\usepackage{graphicx}                   
\usepackage{amssymb}                    
\usepackage{color}                      
\usepackage{breakurl}                   
\usepackage{nicefrac}

\usepackage{rotating}
\usepackage{multirow}

\usepackage{hyperref}
\hypersetup{colorlinks=true,linkcolor=blue,linkcolor=blue,citecolor=blue,urlcolor=blue}

\newcommand{\aap}{    {\it Astron. Astrophys.}}

\newcommand{\apj}{    {\it Astrophys. J.}}
\newcommand{\apjl}{   {\it Astrophys. J. Lett.}}

\newcommand{\nat}{    {\it Nature}}

\newcommand{\solphys}{{\it Solar Phys.}}

\newcommand{\alfven}{Alfv\'en}
\newcommand{\radius}{1\,Mm}
\newcommand{\aspect}{50:1}

\begin{document}
\begin{article}
\begin{opening}
\title{Multi-Stranded Coronal Loops: Quantifying Strand Number and Heating Frequency from Simulated \textit{Solar Dynamics Observatory} (SDO) \textit{Atmospheric Imaging Assembly} (AIA) Observations}
%
\author[addressref={1},corref,email={tomwilliamsphd@gmail.com}]{\inits{T.}\fnm{Thomas}\lnm{~Williams}\orcid{0000-0002-2006-6096}}
\author[addressref={1}]{\inits{R.~W.}\fnm{Robert~W.}\lnm{~Walsh}\orcid{0000-0002-1025-9863}}
\author[addressref={2}]{\inits{S.}\fnm{Stephane}\lnm{~R\'egnier}\orcid{0000-0001-8954-4183}}
\author[addressref={3}]{\inits{C.~D.}\fnm{Craig~D.}\lnm{~Johnston}\orcid{0000-0003-4023-9887}}

%
\runningauthor{T. Williams et al.}
\runningtitle{Quantifying Strand Number and Heating Frequency}

\address[id={1}]{Jeremiah Horrocks Institute, University of Central Lancashire, Preston, UK, PR1 2HE, UK}
\address[id={2}]{Department of Mathematics, Physics, and Electrical Engineering, Northumbria University, Newcastle upon Tyne, United Kingdom, NE2 1XE, UK}
\address[id={3}]{School of Mathematics and Statistics, University of St Andrews, Fife, United Kingdom, KY16 9SS, UK}
\begin{abstract}
Coronal loops form the basic building blocks of the magnetically closed solar corona yet much is still to be determined concerning their possible fine-scale structuring and the rate of heat deposition within them. Using an improved multi-stranded loop model to better approximate the numerically challenging transition region, this paper examines synthetic NASA \textit{Solar Dynamics Observatory}'s (SDO) \textit{Atmospheric Imaging Assembly} (AIA) emission simulated in response to a series of prescribed spatially and temporally random, impulsive and localised heating events across numerous sub-loop elements with a strong weighting towards the base of the structure; the nanoflare heating scenario. The total number of strands and nanoflare repetition times are varied systematically in such a way that the total energy content remains approximately constant across all the cases analysed. Repeated time lag detection during an emission time series provides a good approximation for the nanoflare repetition time for low-frequency heating. Furthermore, using a combination of AIA 171/193 and 193/211 channel ratios in combination with spectroscopic determination of the standard deviation of the loop apex temperature over several hours alongside simulations from the outlined multi-stranded loop model, it is demonstrated that both the imposed heating rate and number of strands can be realised.
\end{abstract}

%

\end{opening}

\section{Introduction}
Understanding the basic physical properties of coronal loops is essential if we are to determine the dominant energy transport mechanisms operating from the solar interior through the chromosphere and into the corona. Observed across all wavelengths, loops are the intrinsic building blocks of the magnetically closed solar atmosphere. These plasma loops range in temperature from a few 10,000\,K up to 10 MK, can channel plasma flow up to 100\,km\,s\textsuperscript{-1} \citep{fredvik02} and show significant transverse oscillations following large impulsive events \citep{nakariakov19}. But do current observational capabilities indicate that we can resolve and definitively determine the inherent nature of these ubiquitous structures?

The launch of the \textit{Atmospheric Imaging Assembly} (AIA: \citealp[ 0.6\ensuremath{^{\prime\prime}} pixel]{lemen12}) onboard NASA's \textit{Solar Dynamics Observatory} (SDO) has provided 24/7 observations of the whole Sun over a wide EUV wavelength range. Coupled to the \textit{Extreme Ultraviolet Imaging Spectrometer} (EIS: \citealp{culhane07}) on \textit{Hinode}, the resulting work on solar atmospheric loops is extensive (e.g. \citealp[etc.]{mcintosh09,brooks12,reale14,xie17}). Subsequently, the superior resolution images from NASA’s sounding rocket \textit{High-resolution Coronal imager} (Hi-C: \citealp[0.1\ensuremath{^{\prime\prime}} pixel]{kobayashi14}) has led to observational assertions of the detection of magnetic braiding \citep{cirtain13}, nanoflares in inter-moss flares \citep{winebarger13}, and the determination of plasma loop parameters \citep{brooks13,peter13,scullion14,aschwanden17,williams20a}. It is still not fully settled whether the loop structures we observe with our current capabilities are actually spatially resolved or if they may be comprised of several individually isolated sub-resolution strands \citep{peter13,xie17,williams20b}.

When examining the make-up of coronal loops, one important factor for consideration is whether the observed loop structure is isothermal or multi-thermal along the line of sight. Differential emission measure (DEM) analysis of loop cross-sections have demonstrated both (\citealp{warren08,brooks12}). Also consider the work by \citet{schmelz16} where the authors create a coronal-loop inventory of 99 loop observations, finding 61 required multi-thermal loop models to reproduce the observations, another 28 being nearly or effectively isothermal with another 10 firmly isothermal. Thus, when determining the temperature at a specific location in a loop (at the apex, say), if this yields a multi-thermal cross-section, then there is the argument that there could be either other plasma along the line of sight or sub-resolution elements existing in the loop that contribute a range of temperatures at that loop location. On the other hand, if there is a single dominant temperature recorded, then either the structure at that point is resolved (is monolithic) and hence at that temperature or if there are strands, their temperatures at the observed location are all very similar -- they appear coherently. Determining a resolved fundamental spatial scale or the presence of sub-elements within a loop is an important step in addressing, for example, how coronal plasma is possibly being heated.

One of the favoured approaches attempting to explain the heating of coronal loop structures is the nanoflare model \citep{parker83,parker88}. Here coronal plasma is heated to million degree temperatures due to the occurrence of numerous, small-scale, localised bursts of energy somewhere within the overall loop envelope. The energy release could be between 
 $10^{23}$\,--\,$10^{27}$\,erg, with possible differing mechanisms of the underlying physics being magnetic reconnection \citep{parker88} or magnetohydrodynamic wave dissipation \citep{satoshi04}. \citet{antolin08} demonstrate a link between the power-law index and the operating heating mechanism in a loop using two 1.5D models; one of \alfven\ wave dissipation and one of random energy bursts to simulate nanoflares. From this they obtain different coronae for the two models, which can be seen in simulated Hinode/XRT flow patterns. It is expected the dissipation of \alfven\ waves would lead to frequent, short-lived heating events along individual magnetic field lines \citep{asgari12}. However, \citet{antolin20} have recently observed the presence of ``nanojets'' in coronal loops. These ``nanojets'' are postulated to be a signature of coronal heating caused by localised magnetic-reconnection activity and the results strongly suggests nanoflare-storms arise due to curved magnetic-field lines reconnecting at small angles. Investigating the heating frequency of individual strands in the corona could further reveal the nature of nanoflare events, that is, whether they are caused exclusively by small-scale magnetic reconnection events or whether they may also be the result of \alfven\ waves. 

Additionally, observations from different wavelength ranges provide strong constraints on the heating mechanism of loop structures \citep{cargill93,cargill94,cargillklimchuk97,cargillklimchuk04}. Following this, \citet{mendoza02} present numerical calculations detailing the response of coronal plasma to periodic, random micro-scale heating pulses within a magnetic loop. The results show successive energy bursts can maintain the coronal plasma to typical coronal temperatures, as well as providing good qualitative agreement with \textit{Transition Region and Coronal Explorer} (TRACE) loops observations \citep{aschwanden01}. More recently \citet{cargill15} demonstrated that the time between nanoflare events on individual magnetic strands should be between 500\,--\,2000\,seconds, where the energy release may be as small as a few $10^{23}$\,erg.

One numerical approach that provides a quick and accurate answer to the coronal response of a loop to heating is the use of zero-dimensional (0D) field-aligned hydrodynamical models (EBTEL: \citealp{klimchuk08,cargill12a,cargill12b,cargill15}). As each loop is modelled as a single grid point, 0D models allow the coronal-loop evolution to be simulated very efficiently with minimal computational cost, whilst still being capable of obtaining time-dependent, spatially averaged loop quantities that are comparable to fully resolved 1D simulations. Consequently, this method for simulating the evolution of coronal loops has proved popular \citep{viall11,cargill15,barnes16}.

However, 1D simulations are still greatly useful, particularly those which model a multi-stranded structure as opposed to monolithic loops. For example, \citet{price15} forward model the synthetic profiles of a well-documented \textit{Hinode}/EIS structure \citep{mcintosh09} using various numbers of sub-element strands to try and deduce the nature of the outflows observed. Their results support a scenario whereby long loops formed of multiple strands undergo periodic heating and cooling. Similarly, \citet{susino13} demonstrate that whilst 1D multi-stranded models can predict observed values derived from TRACE and AIA filter ratios, they cannot explain all of the characteristics of warm over-dense loops as such models assume spatial interactions between individual strands can be neglected, which may not always be the case \citep{hood16,reid18,reid20}.

Building upon and improving the computational approach of the multi-\,\,\, stranded loop code from \citet{sarkarwalsh08,sarkarwalsh09}, this article uses synthesised emission as would be observed by AIA to investigate whether over a long time series of data (six hours) it is possible to recover an imposed heating frequency as well as any information about potential coronal loop sub-elements that are beyond the spatial resolving power of the instrument -- such as the recent high-resolution strand observations by \citet{williams20a,williams20b}. Section 2 details the updated numerical approach that now incorporates the unresolved transition region (UTR) method developed by \citet{johnstonhood17a,johnstonhood17b}. In Section 3, an analysis of the loop apex temperature evolution from 18 different loop configurations is detailed. The data is employed to recreate synthetic AIA emission in Section 4 where the resulting light curves are examined using methods such as emission-line ratio comparisons and time-lag analysis \citep{viall11}.The findings of this study and their implications on the determination of heating rate and strand number are discussed in Section 5.

\section{Improved Multi-Stranded Loop Model}
The numerical code employed in this article is written in FORTRAN and is based on the Lagrangian remap method (LareXd) developed by \citet{arber01}, which has been modified for multi-stranded coronal loops \citep{sarkarwalsh08,sarkarwalsh09}. A unique feature of this code is that each individual sub-element strand of a loop is an individual, independent simulation, all of which are then amalgamated in post-processing to form a single coherent coronal loop. The code is managed in such a way that the loop/strand parameters (such as nanoflare energy, loop/strand length and radius, etc) can be edited without the need for recompiling the model, allowing for multiple instances/scenarios to run concurrently. In this article, we continue the development of the multi-stranded hydrodynamic loop (MSHDL) simulation by treating the transition region as an unresolved discontinuity across which energy is conserved by imposing a jump condition: the Unresolved Transition Region (UTR) method from \citet{johnstonhood17a,johnstonhood17b}.

As will be discussed in this section, adopting the UTR method means the resolution of the MSHDL simulations can be reduced with minimal impact on the ``observables'' compared to other loop models \citep{johnstonhood17a,johnstonhood17b}, whilst also decreasing the computational time of each strand; an important factor when each loop is comprised of numerous strands. Similar approaches have been also been adopted in previous studies (for example, see \citealp{mikic13,johnston19,johnstonbradshaw19,vandamme20}).

\subsection{Initial Model Setup}
The plasma-$\beta$ is the ratio of thermal pressure to magnetic pressure within a given plasma. In the corona, and structures such as coronal loops, the $\beta < 1$. This means the plasma is ``tied-in'' to the magnetic field, and subsequently, the shape, size, and orientation of the plasma structures are determined by the field lines. In addition to a low plasma-$\beta$, the corona is also a highly conducting medium. These properties allow for the assumption that the plasma dynamics occur along magnetic field lines, with negligible feedback between the magnetic field and thermodynamic parameters. As is discussed by \citet{sarkarwalsh08,sarkarwalsh09}, these physical attributes support the use of a 1D hydrodynamic code, which neglects thermal conduction across field lines when modelling entire loop-like structures.

With this in mind, here a loop is modelled as an amalgamation of a collection of individual, narrower, sub-loop element, which here are called ``strands". Each strand is assumed to have the same cross-sectional radius and total length. The MSHDL code is a 1D hydrodynamic adaptation of LareXd \citep{arber01} that simulates each of these strands individually, which are then combined in post-processing to form a single loop structure. These loops can then be compared with observed loop structures that may appear as monolithic in nature but may actually comprise several-to-hundreds of structures whose spatial scales are below current instrument capabilities.

As aforementioned, the modelled loop is a compilation of single strands [$i$], each obeying the standard equations for mass, momentum, and energy conservation in curvilinear abscissa along a semi-circular loop:

\begin{equation}\label{eq:hd_mass}
\frac{\mathrm{D}\rho_i}{\mathrm{D}t} =-\rho_i \frac{\partial v_i}{\partial s},
\end{equation}

\begin{equation}\label{eq:hd_momentum}
\rho_i \frac{\mathrm{D}v_i}{\mathrm{D}t} = -\frac{\partial p_i}{\partial s} + \rho_i g + \rho_i v\frac{\partial^2v_i}{\partial s^2},
\end{equation}

\begin{equation}\label{eq:hd_energy}
\frac{\rho_i^\gamma}{\gamma-1}\frac{\mathrm{D}}{\mathrm{D}t}\left(\frac{p_i}{\rho_i^\gamma}\right) = \frac{\partial}{\partial s}\left(\kappa\frac{\partial T_i}{\partial s}\right) - n_i^2Q\left(T_i\right) + H_i\left(s,t\right),
\end{equation}
where
\begin{equation}\label{eq:p}
p_i = \frac{R}{\tilde{\mu}}\rho_i T_i,
\end{equation}
and
\begin{equation}\label{eq:Dt}
\frac{\mathrm{D}}{\mathrm{D}t} \equiv \frac{\partial}{\partial t} + v_i \frac{\partial}{\partial s}.
\end{equation}
Following standard notation, $\rho_i$, $p_i$, and $T_i$, are the density, thermal pressure, and temperature of strand $i$. The velocity, [$v_i$] is along the curvilinear abscissa [$s$] of the strand. The adiabatic index, and mean molecular mass are given by $\gamma=\nicefrac{5}{3}$ and $\tilde{\mu}=0.6 \textrm{\,mol}^{-1}$, respectively. Gravity is assumed to be constant and is taken to be the surface gravity: $g = 2.74 \times 10^4 \textrm{\,cm~s}^{-2}$. The number density [$n_i$] is defined as $10^{21}$\,[c$\textrm{m}^{-3}$] $\rho_i$\,[kg\,$\textrm{m}^{-3}$]. The conductivity of the plasma in the direction of $s$ is denoted by $\kappa = 9.2\times 10^{-7}T^{5/2}\,\textrm{\,erg\,s}^{-1}\,\textrm{\,cm}^{–1}\,\textrm{\,K}^{–1}$. $R = 8.3 \times 10^7\,\textrm{\,erg\,mol}^{-1}\,\textrm{K}^{-1}$ is the molecular gas constant, and $Q(T_i)$ is the optically thin radiative loss function \citep{rosner78}. The remaining term [$H_i\left(s,t\right)$] is our heating input for the individual strands, which is discussed briefly in Section \ref{sec:heating} and presented in more detail by \citet{sarkarwalsh08,sarkarwalsh09}.

For the simulated loop, a coronal length of $L=100$\,Mm is adopted. The loop foot-points are at $\nicefrac{-L}{2}$ and $\nicefrac{+L}{2}$, with the loop apex situated at $s = 0$. The loop radius is \radius\  giving the loop an aspect ratio of \aspect, which is consistent with high-resolution observations for loops longer than 50\,Mm \citep{peter13}. The area of each strand is approximated as $A_{\mathrm{strand}} = A_{\mathrm{loop}}/N_{\mathrm{strands}}$, where $A$ is cross-sectional area, and $N_{\mathrm{strands}}$ is the total number of strands forming the loop.

The boundary conditions are prescribed as
\begin{equation}\label{eq:MSHDL_BC1}
T\left(-L/2,t\right) = T\left(+L/2,t\right) = T_{\mathrm{ch}} = 10^4\textrm{\,K},
\end{equation}
and
\begin{equation}\label{eq:MSHDL_BC2}
p\left(-L/2,t\right) = p\left(+L/2,t\right) = p_{\mathrm{ch}} = 0.315\textrm{\,Pa},
\end{equation}
at the loop footpoints at the base of the chromosphere, which is 5\,Mm deep. Note that $T_{ch}$ and $p_{ch}$ are the chromospheric temperature and pressure.

To define the temperature of the loop from the amalgamated strands, the emission-measure temperature is calculated, as is described by \citet{sarkarwalsh09}, by
\begin{equation}\label{eq:em_temp}
    T = \frac{\sum\limits_{i} \rho_i^2\left(s,t\right) T_i\left(s,t\right)}{\sum\limits_{i} \rho_i^2\left(s,t\right)} .
\end{equation}

\begin{figure*}[htp!]
\centerline{\includegraphics[width=0.85\textwidth]{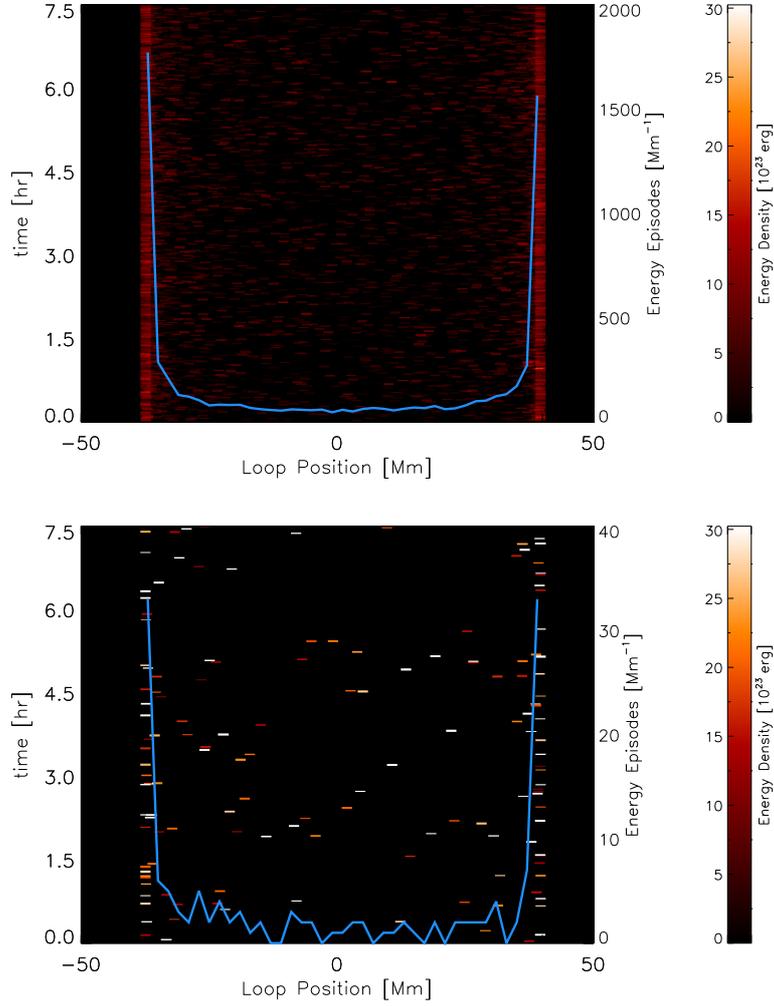}}
\caption{The \textit{top} (\textit{bottom}) panel shows the spatio-temporal plot indicating the duration, position, and magnitude of each energy episode within a loop consisting of 64 (8) strands, which is subjected to localised heating events every 250\,seconds (1500\,seconds) with minimum flare energy $2.5\times10^{23}$\,erg ($1.2\times10^{25}$\,erg). The magnitudes of the energy episodes are shown on the same colour table, whilst the blue lines indicate the nanoflare number densities as a function of loop position for the two scenarios. The total energy deposit of each loop is equivalent and hence together, the two panels provide a comparison of the heating distribution between a loop that is heated by many small-magnitude energy episodes (\textit{top}) and a loop that is subjected to a few energy episodes that have a large magnitude (\textit{bottom}).}
\label{fig:nanoflares}
\end{figure*}

\subsubsection{Mimicking Nanoflare Heating}\label{sec:heating}
The heating is prescribed on each strand as a series of small, localised, energy-release episodes akin to the nature of nanoflares. The nature of these small energy bursts are governed by a random-number generator, such that their spatial and temporal distributions are unique and distinct for each strand of the loop. This heating approach is described fully by \citet{sarkarwalsh09}, and so they are just restated briefly here.

Each heating episode has a minimum energy threshold, which may be set by the user -- although a typical energy would be $10^{24}\,\textrm{erg}$. During the simulation, each strand is subject to a user-specified number of energy episodes whose spatial deposition is random but is weighted such that the energy release predominantly occurs near the base of the strands between loop positions $\pm$35\,--\,39\,Mm (Figure\,\ref{fig:nanoflares}). The length-scale of each heating event is fixed at 2\,Mm whilst the timescale is randomly distributed between 50\,--\,150\,seconds.

\subsection{Modelling the Transition Region}
In this article we consider the transition region to be a thin boundary layer between the chromosphere and corona where there is a rapid increase (decrease) in temperature (density). In coronal loop simulations, the ability to accurately resolve the behaviour of this boundary layer is important. For example, \citet{bradshawcargill13} demonstrate that inadequately reproducing the transition region leads to artificially low coronal densities. In this article, the approach outlined by \citet{johnstonhood17a,johnstonhood17b} is adopted in the MSHDL code by treating the lower transition region as a discontinuity that responds to changing conditions -- such as mass flow and plasma heating -- through the imposition of a jump condition derived from an integrated form of energy conservation. This method has been shown to capture accurately the coronal density evolution of fully resolved 1D models (e.g. \citealp{bradshawcargill13}), whilst also being computationally significantly more efficient.

Here, the key points for modelling the transition region using the unresolved transition region (UTR) method are briefly outlined, as per Sections 2 and 3 of \citet{johnstonhood17a,johnstonhood17b}. The base of the UTR is defined as the location where the temperature first reaches, or drops below the chromospheric temperature ($10^4$\,K) when moving from the strand apex towards the chromosphere along the curvilinear abscissa [$s$]. Below the base of the UTR, the temperature is held fixed at the chromospheric temperature -- $10^4$\,K.

The top of the UTR is defined as the final location at which the following criterion is satisfied when travelling away from the apex along $s$:
\begin{equation}\label{eq:UTR_top}
\frac{L_R}{L_T} \leq \delta < 1,
\end{equation}
where $L_R$ and $L_T$ are the spatial and temperature length scales of the simulation as defined by \citet{johnstonhood17a}. In order to ensure that multiple gridpoints are maintained across the temperature length scale at this region, $\delta=0.15$ is selected. These two definitions allow for easy identification of the UTR across all time steps of the simulation. 

Having identified the UTR, the second part of the UTR method is the imposition of the jump condition at the top of the UTR, which takes the form of a local velocity correction which conserves the total energy in the UTR and is imposed at each time step. The jump condition is reproduced here from Equation\,12 of \citet{johnstonhood17a}  
\begin{equation}\label{eq:jump}
    \frac{\gamma}{\gamma-1}p_0v_0 + \frac{1}{2}\rho_0v_0^3 + \rho_0\Phi_0v_0 = l_{\mathrm{UTR}}\bar H - (R_{\mathrm{UTR}}+F_{c,0}),
\end{equation}
where the left-hand terms denote the flux for enthalpy, kinetic energy, and gravitational potential energy, respectively. As for the right-hand, the terms describe the averaged volumetric heating rate per unit cross-sectional area, the integrated radiative losses in the UTR [$R_{UTR}$] and heat flux [$F_c$]. $\Phi$ is the gravitational potential, $\bar H$ is the spatially averaged volumetric heating rate, and $l_{\mathrm{UTR}}$ is the length of the UTR. The subscript 0 in Equation\,\ref{eq:jump} denotes quantities measured at the top of the UTR. The full details on the UTR method were given by \citet{johnstonhood17a}. 

\subsubsection{Comparison of Two Numerical Approaches}\label{sec:mslutrmshdl}
\begin{figure*}[htbp!]
\centerline{\includegraphics[width=0.925\textwidth]{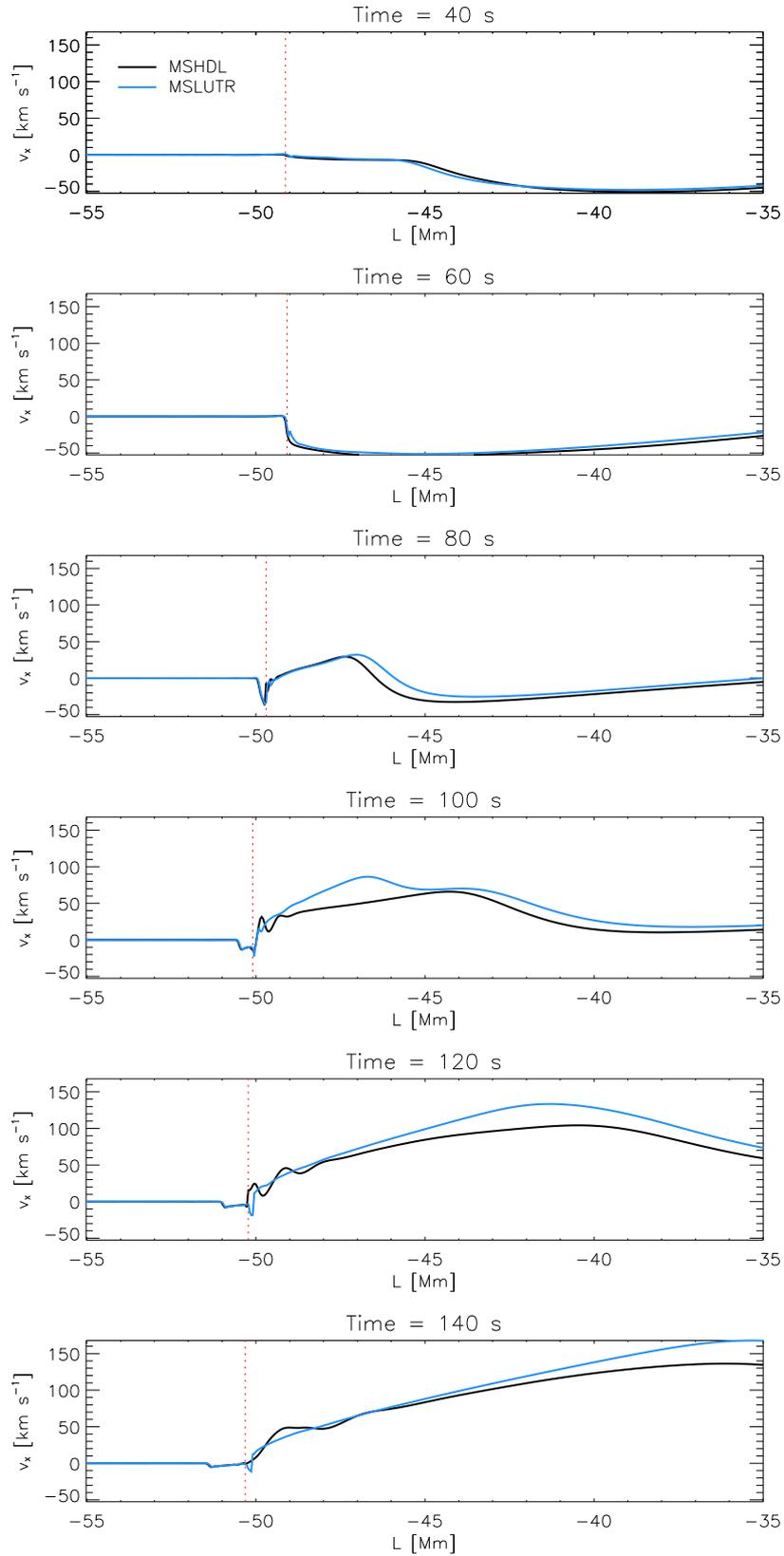}}
\caption{Single-strand velocity profile comparison of the previous MSHDL code (black) vs. new MSLUTR method (blue) for a single burst event near the $-L/2$ footpoint in 20\,second intervals. The base of the transition region is denoted by the vertical dashed-red line. Only one footpoint is shown so that the differences between the models at the transition region can be seen more clearly.}
\label{fig:mshd_compare}
\end{figure*}

Here, a comparison is made between the two versions of the multi-stranded loops code; MSHDL \citep{sarkarwalsh08,sarkarwalsh09} and the updated code, MSLUTR (Multi-Stranded Loops with Unresolved Transition Regions) employing the established UTR jump condition \citep{johnstonhood17a,johnstonhood17b}. For this comparison, a single localised release of energy is modelled at position $s=-23$\,Mm depositing a total energy of $2.3\times10^{24}$\,erg over a length of 2\,Mm for a duration of 1218\,seconds. The temporal evolution of this energy burst near the $-L/2$ footpoint is shown for the two codes at 20\,second intervals in Figure\,\ref{fig:mshd_compare}.

The deposited energy drives a plasma flow that propagates down the strand leg until the conduction front reaches the transition region. Initially, the discrepancies between the MSHDL and MSLUTR are negligible. This can be attributed to the dynamics being set by the direct in-situ heating for the first 60\,seconds. However, once the enhanced downward heat flux reaches the transition region at 60\,seconds notable differences start to arise. This is the start of the evaporation phase and by 80\,seconds the leading front of the flow in MSLUTR is $\approx1$\,Mm ahead of the corresponding flow in MSHDL.

By 100\,seconds it can be seen that MSHDL is underestimating the evaporative upflow compared to MSLUTR. The largest difference in the velocity profiles between the two regimes occurs at $\approx-47$\,Mm;  where MSLUTR velocity  $\approx85$\,km\,s\textsuperscript{-1} vs. $\approx50$\,km\,s\textsuperscript{-1} for MSHDL.

In addition to MSHDL underestimating the coronal velocity, under-resolving the lower transition region also results in spurious oscillations propagating upwards from the chromosphere into the corona. These oscillations can be seen clearly at times \textit{t} = 80, 100, 120, and 140\,seconds. In contrast, these oscillations are significantly less prominent for MSLUTR but a small perturbation can still be seen to the right of the base of the transition region between 80\,seconds\,--\,140\,seconds (red line, Figure\,\ref{fig:mshd_compare}). However, unlike MSHDL, the perturbation is contained within the UTR and does not propagate upwards into the coronal part of the structure. Furthermore, as is evidenced in previous work, adopting the UTR method provides comparative results with other fully resolved 1D simulations of impulsive heating \citep{johnstonhood17a,johnstonhood17b} and the development of thermal non-equilibrium \citep{johnston19}.

Whilst Figure\,\ref{fig:mshd_compare} highlights the improvements to modelling a single strand, one of the main benefits to using this code over other numerical schemes is the fact that it investigates the multi-stranded nature of loop structures. Thus, it is not enough for improved accuracy on a single strand during a single impulsive event if the run-time and/or spatial resolution required to achieve this greater accuracy makes multi-strand analysis impractical. To this end, a maximum velocity convergence test is performed using the aforementioned scenario where the number of gridpoints are varied using the MSLUTR code (Table\,\ref{tab:converge}). From this convergence test, it is deduced that a simulation resolution of \textit{nx} = 2,000 is sufficiently accurate with maximum coronal velocities within $\approx4.5$\,\% of an over-resolved simulation (\textit{nx} = 15,000). This choice allows for improved accuracy compared to the MSHDL code whilst allowing the run-time to remain sufficiently manageable for potentially hundreds of strands to be simulated in a timely manner.
\begin{table}[!ht]
\caption{MSLUTR velocity convergence test relative to \textit{nx} = 15,000 simulation.}
\label{tab:converge}
\begin{tabular}{c c c}
\hline 
Resolution [\textit{nx}] & $-L/2$ Footpoint Difference [\%] & Coronal Difference [\%]\\    
 \hline
 500  & 8.77 & 13.99 \\
 1000 & 5.90 & 8.27 \\
 2000 & 5.95 & 4.52 \\
 5000 & 4.62 & 2.24 \\
 7500 & 1.90 & 0.72 \\
 \hline
\end{tabular}
\end{table}

\section{Multi-Stranded Analysis}\label{sec:multistrand}
\begin{table}
\caption{Loop apex information for the 18 different setups examined.}
\label{tab:param_scan}
\begin{tabular}{c c c c c c c c}
\hline
\multirow{6}*{}{No. of} & {Repetition} & {Frequency} & {Min. Energy} & 
      \multicolumn{2}{c}{Temperature [MK]} &
      \multicolumn{2}{c}{Speed [km\,s\textsuperscript{-1}]} \\
Strands&Time [s]&[$10^{-4}$\,Hz]&[$\times10^{24}$ erg]&Max:&Avg:&Max:&Avg:\\
\hline
8 & 250 & 40.00 & 2.00 & 3.81 & 2.71 & 162 & 42\\
8 & 500 & 20.00 &  4.00 & 4.68 & 2.76 & 373 & 55\\
8 & 750 & 13.33 &  6.00 & 5.11 & 2.74 & 310 & 56\\
8 & 1000 & 10.00 &  8.00 & 6.26 & 2.78 & 390 & 54\\
8 & 1250 & 8.00 &  10.00 & 6.04 & 2.86 & 562 & 46\\
8 & 1500 & 6.67 &  12.00 & 6.21 & 2.79 & 800 & 59\\
\hline
16 & 250 & 40.00 & 1.00 & 3.40 & 2.71 & 252 & 32\\
16 & 500 & 20.00 & 2.00 & 3.94 & 2.81 & 310 & 54\\
16 & 750 & 13.33 & 3.00 & 3.87 & 2.88 & 378 & 49\\
16 & 1000 & 10.00 & 4.00 & 3.78 & 2.78 & 458 & 48\\
16 & 1250 & 8.00 & 5.00 & 4.52 & 2.77 & 718 & 49\\
16 & 1500 & 6.67 & 6.00 & 4.32 & 2.77 & 534 & 51\\
\hline
64 & 250 & 40.00 & 0.25 & 3.47 & 2.76 & 159 & 41\\
64 & 500 & 20.00 & 0.50 & 3.71 & 2.82 & 262 & 57\\
64 & 750 & 13.33 & 0.75 & 4.05 & 2.90 & 389 & 52\\
64 & 1000 & 10.00 & 1.00 &  4.30 & 2.94 & 379 & 51\\
64 & 1250 & 8.00 & 1.25 & 4.93 & 2.96 & 319 & 44\\
64 & 1500 & 6.67 & 1.50 & 4.32 & 2.96 & 382 & 56\\
\hline
\end{tabular}
\end{table}

This section provides results from the improved version of the multi-stranded loop code MSLUTR. The model loop is configured to be 100\,Mm long with a radius of 1\,Mm and initial chromospheric and coronal temperatures of 10\textsuperscript{4}\,K and 10\textsuperscript{6}\,K, respectively. The number of strands and energy release episodes (i.e. nanoflares) the loop is subjected to varies in such a way that each loop configuration has approximately the same total amount of energy deposited (mean: $1.10\times10^{28}$\,erg, $\sigma$: $0.04\times10^{28}$\,erg) during the simulation run, i.e. the greater the number of nanoflares and/or strands within a loop, the smaller the magnitude of energy release per nanoflare that occurs. Since the total energy deposition is approximately the same across all the simulations, direct comparisons can be made between the different heating frequencies investigated. The loop is simulated for a total time of 7.5\,hours, but the first 1.5\,hours are ignored. The reasoning behind this is that each strand requires several acoustic timescales to elapse due to 
i) the sudden initial injection of energy generating large-scale perturbations to damp throughout the system and 
ii) several heating events needing to occur to allow a quasi-steady equilibrium to be reached. 
Ignoring the first 1.5\,hours is an arbitrary value that is sufficiently long to account for these hence allowing 6\,hours to be analysed from each generated data-set.

\begin{figure*}[htp!]
\centerline{\includegraphics[width=\textwidth]{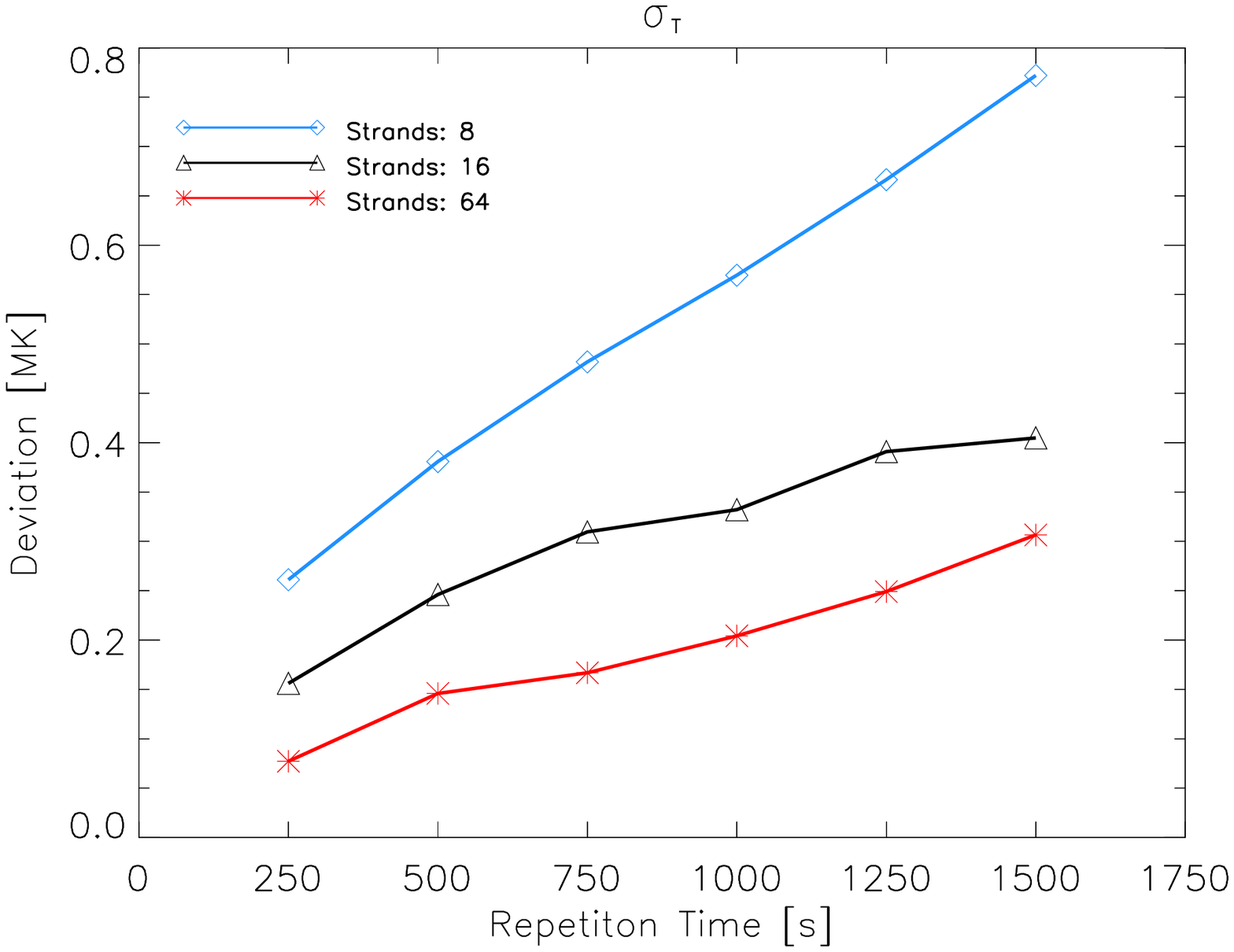}}
\caption{The average standard deviation in temperature at the loop apex for all simulations plotted against nanoflare repetition rate.}
\label{fig:std_dev}
\end{figure*}
In the literature regarding nanoflare heating frequencies, it is common to use the repetition time when discussing frequency due to the units being more meaningful, and as such we adopt this nomenclature for the remainder of this study (Table\,\ref{tab:param_scan} also lists the nanoflare frequency in Hz for completeness). For example, \citet{cargill15} argue that the repetition time of nanoflares along an individual strand to be in the range of 500\,--\,2000\,seconds.  With this in mind, nanoflare repetition times are explored that correspond to low-frequency heating ($\approx1500$\,seconds; termed LFH), intermediate-low frequency heating ($\approx1250$\,seconds; ILFH) intermediate frequency heating 
\linebreak ($\approx1000$\,seconds; IFH), intermediate-high frequency heating ($\approx1500$\,seconds; IHFH), high frequency heating ($\approx500$\,seconds; HFH), and ultra-high frequency heating ($\approx250$\,s; UHFH). Three sets of loops with increasing numbers of strands are explored -- 8, 16, and 64 elements. The details of the resulting eighteen loop configurations examined are shown in Table\,\ref{tab:param_scan}. For the configurations tested in this study, the mean apex temperatures of the loop (as calculated using Equation\,\ref{eq:em_temp}) is in the narrow range of between 2.71\,--\,2.96\,MK, a consequence of the injected energy being approximately equal for each loop configuration (mean: $1.10\times 10^{28}$\,erg, $\sigma$: $0.04\times 10^{28}$\,erg) despite the varying number of strands and nanoflares taking place.

In Figure\,\ref{fig:std_dev} the standard deviation of loop apex temperature compared to their mean temperatures (Table\,\ref{tab:param_scan}) is shown for the 6\,hour window displayed as a function of nanoflare repetition time. To understand these plots, consider a loop with a fixed number of strands; as the time between events increases (the repetition time increases), the magnitude of each individual nanoflare must also increase to maintain a fixed total energy deposition budget across the entire simulation. Consequently, more sparse events of larger size in the strands will create larger variations in the responding physical parameters of the loop. In contrast, many strands with a shorter delay between nanoflares have smaller energy release magnitudes and subsequently alter the physical parameters of the loop significantly less -- hence a significantly reduced standard deviation.

\section{Synthesised AIA EUV Emission}\label{sec:synthAIA}
 Synthetic AIA emission is calculated for the 18 loop configurations shown in Table\,\ref{tab:param_scan} by the following equation outlined in \citet{sarkarwalsh09}:
\begin{equation}\label{eq:aia_emission}
EM_{\mathrm{AIA}} = \sum_{i}^{N_{\mathrm{strands}}}\rho_i^2 C\left(T_i\left(s\right)\right)\; \mathrm{d}s,
\end{equation}
where $i=1,2,3,...\;N_{\mathrm{strands}}$ is the strand number. $EM_{\mathrm{AIA}}$ is the synthetic AIA emission, $\rho_i$ is the strand density, and $C\left(T_i\left(s\right)\right)$ is the AIA temperature response function, which is dependent on temperature [$T_i\left(s\right)$] and is measured for all positions $s$ within the loops. The AIA temperature-response functions obtained from CHIANTI v9.0.1 are shown in Figure\,\ref{fig:aia_resp}.

\subsection{Examination of Time Lags Between EUV Line Pairs}\label{sec:TLpairs}
Following the approach of \citet{viall11}, the synthesised AIA emission is employed to examine any possible time-lags between pairs of lines in the resulting emission-line time-series. Here, eight gridpoints at the loop apex from each simulation are convolved to match the resolution of an AIA pixel, for which the AIA emission is then obtained from Equation\,\ref{eq:aia_emission} to produce the synthetic light curves; this is demonstrated in Figures\,\ref{fig:aia_lines} and \ref{fig:aia_lines_2}. The cross-correlation of two light curves that are closest in temperature (e.g. AIA 211 and 193) are computed using the c\_correlate function in IDL \citep{fuller95}. As with \citet{viall11}, a positive (negative) time lag means a peak in emission occurs for the hotter AIA channel before (after) the cooler AIA channel emission peaks.

\begin{figure*}
\centerline{\includegraphics[width=\textwidth]{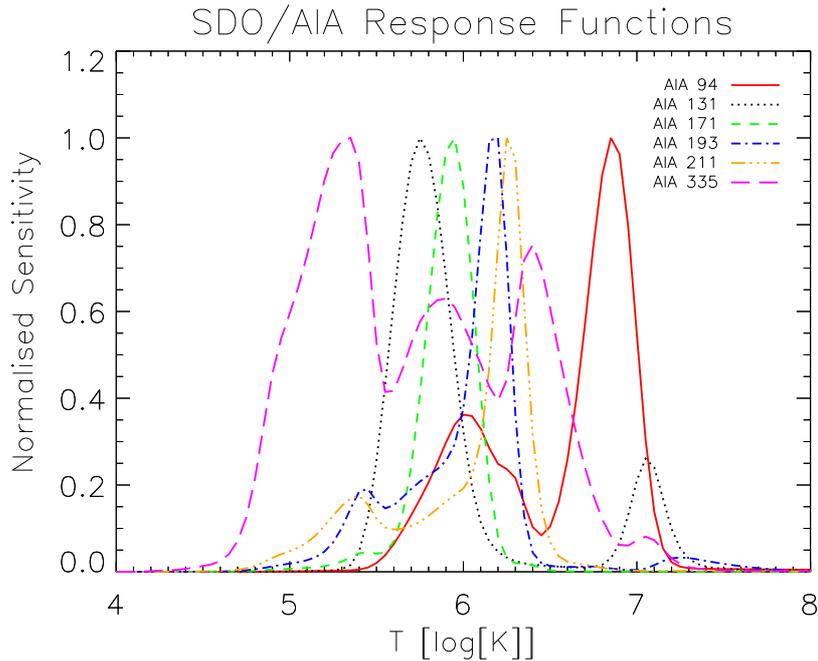}}
\caption{AIA temperature-response functions used to generate synthetic channel temporal profiles for the 18 loop configurations examined.}
\label{fig:aia_resp}
\end{figure*}
\begin{figure*}[!htbp]
\centerline{\includegraphics[width=\textwidth]{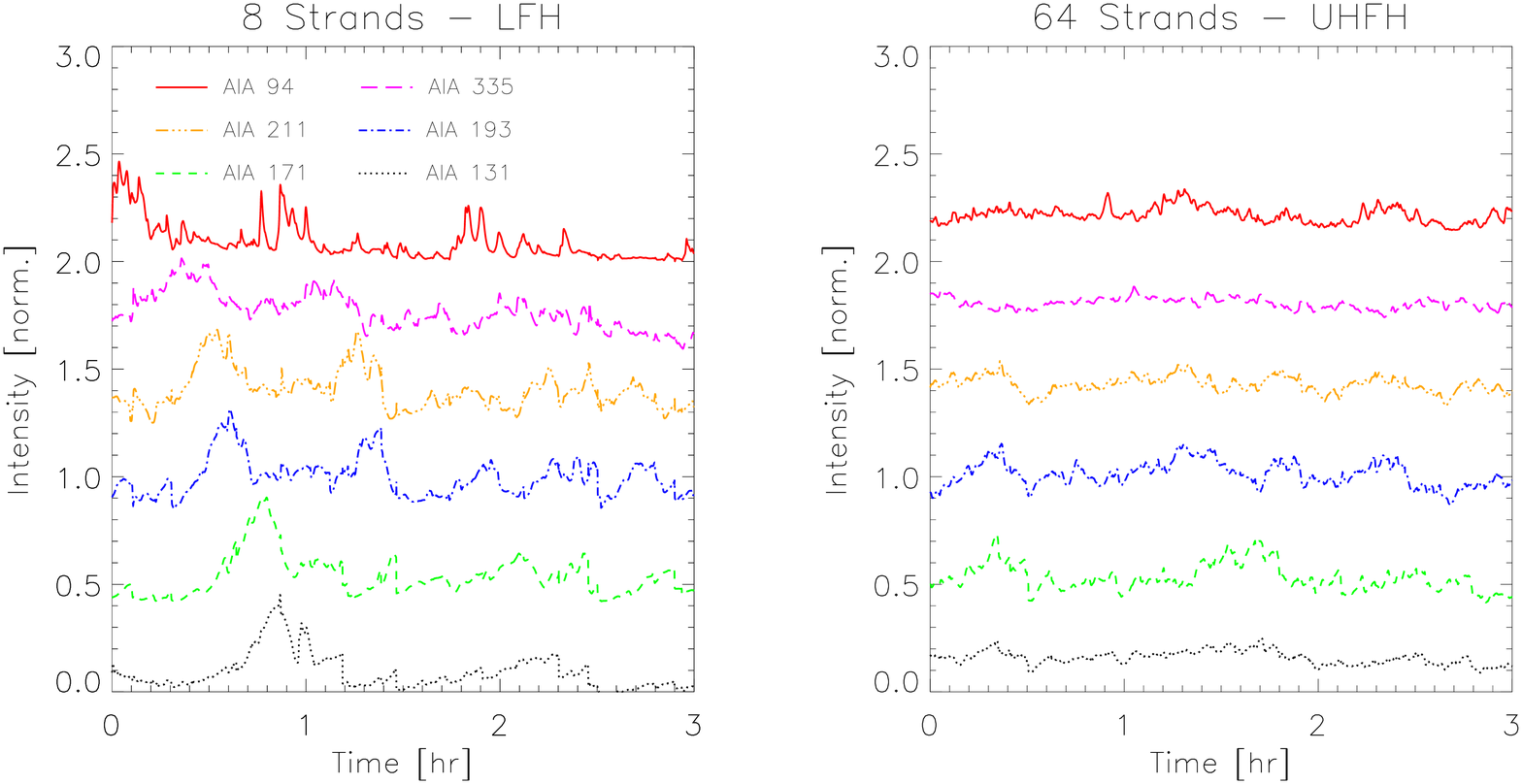}}
\caption{Synthetic AIA channel time profiles over an example three-hour period for the lowest (\textit{left}: 1500-second nanoflare repetition time on 8 strands) and highest (\textit{right}: 250-second nanoflare repetition time on 64 strands) frequency heated loop scenarios in this study. The figure follows the same colour scheme as Figure\,\ref{fig:aia_resp} with AIA 94 (\textit{red}), 335 (\textit{magenta}), 211 (\textit{orange}), 193 (\textit{blue}), 171 (\textit{green}), and 131 (\textit{black}) shown in descending order for the two examples in normalised units with an appropriate offset so that the line profiles can be stacked.}
\label{fig:aia_lines}
\end{figure*}
\begin{figure*}[!htbp]
\centerline{\includegraphics[width=\textwidth]{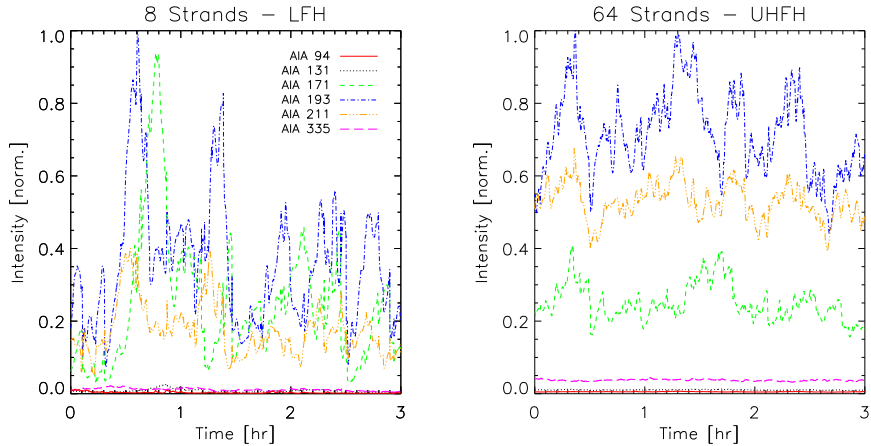}}
\caption{Synthetic AIA channel time profiles over an example three-hour period for the lowest (\textit{left}: 1500-second nanoflare repetition time on 8 strands) and highest (\textit{right}: 250-second nanoflare repetition time on 64 strands) frequency heated loop scenarios in this study. The figure follows the same colour scheme as Figure\,\ref{fig:aia_resp} with AIA 94 (\textit{red}), 335 (\textit{magenta}), 211 (\textit{orange}), 193 (\textit{blue}), 171 (\textit{green}), and 131 (\textit{black}), which have been re-normalised with respect to the AIA 193 channel to show the relative emission of each channel.}
\label{fig:aia_lines_2}
\end{figure*}

Mimicking Figure\,7 of \citet{viall11}, Figure\,\ref{fig:aia_lines} shows an example three-hour time window from the two most extreme cases of heating frequency simulated, as viewed across the AIA channels. The left (right) panel represents the case of a loop with the lowest (highest) heating frequency for 8 (64) strands that are subjected to nanoflare events with a minimum energy release of $1.2\times 10^{25}$\,erg ($2.5\times 10^{23}$\,erg) every $\approx1500$\,seconds ($\approx250$\,seconds). The emission line in Figure\,\ref{fig:aia_lines} is normalised relative to the maximum intensity of each emission line and then offset to produce a stacked line plot. In contrast, Figure\,\ref{fig:aia_lines_2} displays the intensity of each emission line normalised relative to the maximum intensity of the exceptionally bright AIA 193 channel from the three-hour time period under investigation.

In both of the extreme cases presented here, it is possible to follow the plasma evolution across the AIA channels throughout the three-hour period as it heats in response to the nanoflare events and subsequently cools. By examining the resulting time lags of the calculated emission between lines adjacent in peak emission temperature, any possible relationship of this to the input heating frequency can then be explored.

Focusing on the LFH scenario presented in Figure\,\ref{fig:aia_lines} (left) and between $t=0-1$\,hour, a large rise and fall in emission can be seen in AIA 94, that is followed later by a rise and fall in AIA 335 and so on, which, following the nomenclature outlined by \citet{viall16}, results in a positive time lag of $\approx 480$\,seconds between neighbouring AIA channel pairs based on the peak-temperature ordering shown in Figures \ref{fig:aia_resp} and \ref{fig:aia_lines}. This time lag is due to the infrequent, large magnitude energy release from each nanoflare ($\gtrsim1.2\times10^{25}$\,erg) generating large variability in the light curves. Conversely, the UHFH example (Figure\,\ref{fig:aia_lines} right) shows minimal variability in the light curves and subsequently no easily discernible time lag is present between the AIA channels due to the large number of smaller energy nanoflares taking place ($\gtrsim2.5\times10^{23}$\,erg).

\begin{figure*}[!ht]
\centerline{\includegraphics[width=0.75\textwidth]{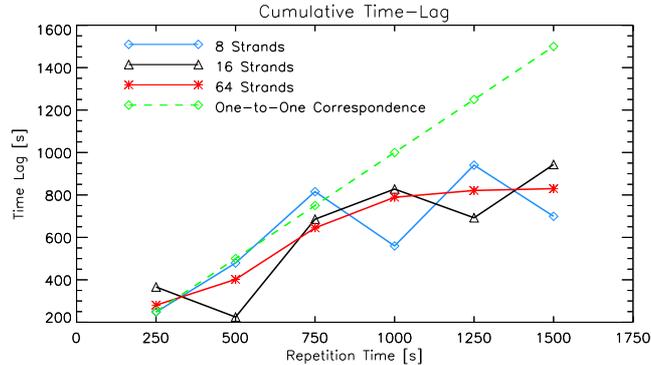}}
\caption{The cumulative time lag for emission detected in AIA 211 to cool and be detectable in the subsequent AIA channels 193 and 171 (as demonstrated for Figure\,\ref{fig:aia_lines}) versus the nanoflare repetition time. All the strand-heating frequency scenarios tested  are shown for loops consisting of 8 (\textit{blue}), 16 (\textit{black}), and 64 (\textit{red}) strands. The \textit{green} dashed line displays where the one-to-one correspondence between the heating frequency and time lag would occur.}
\label{fig:timelags}
\end{figure*}

In Figure\,\ref{fig:aia_lines_2}, the light curves of Figure\,\ref{fig:aia_lines} are re-normalised to the maximum intensity of the AIA 193, which given the peak average loop temperature of $\approx2.7$\,MK (Table\,\ref{tab:param_scan}) is exceptionally bright across all 18 loop configurations analysed in this study. In doing so, it can be seen that AIA 171, 193, and 211 have vastly superior counts compared to the hotter channels (335, 94, and 131). This is largely due to i) the loop parameters being tuned such that the resulting strand temperatures strongly correspond to the peak emission temperatures of the cooler AIA channels and ii) the hotter channels' temperature response functions having lower relative peaks and broader distributions compared to the relatively cooler channels. Additionally, multiple peaks with contributions from cooler temperatures ($\approx1$\,MK) complicate analysis of the hotter emission line intensities \citep{odwyer10}. Considering these factors, the following analysis will purely focus on the emission from the AIA channels 171, 193, and 211.

Figure\,\ref{fig:timelags} provides a means for quantifying the light curves' variability over the six-hour duration for the emission lines analysed. Here, the cross-correlation of the AIA emission line pairs 211--193 and 193--171 are computed and summed together to generate a cumulative time lag to test whether any correlation between time lag and nanoflare repetition time can be identified\footnote{The time lag of 211--171 has also been computed, and for monolithic structures one would expect the two methods to return the same time lag. However, for complex structures consisting of a few to many sub-element strands, the differences in light curves for AIA 211 and AIA 171 can be significant and may result in the c\_correlate function matching peaks/troughs in the light curves resulting from separate, independent events. For this reason, a direct correlation between AIA 211 and AIA 171 is more difficult and, as such, it is found to be better to adopt the intermediate step of analysing the cumulative time lag between 211--193 and 193--171 for this study.}. As would be expected, a general trend can be seen whereby the time lag is shorter for loops whose strands are heated more frequently (smaller repetition time) compared to stands that are heated more infrequently (larger repetition time). For the loops comprised of the most frequently heated strands, the time lag between emission is $\approx 250$\,seconds, which is on the order of the input heating repetition time. As the time between heating events increases, so too does the time lag, although for repetition times of $\gtrsim1000$\,seconds the time lag significantly differs from the nanoflare repetition time. As the heating frequency further decreases towards our lowest frequency examples (1500\,second repetition time), the deviation between repetition time and time lag increases with the time lag plateauing at around the 800\,--\,1000\,second level in all of the multi-stranded scenarios in the case study.

These time lags are a direct consequence of the competition between the fundamental cooling processes operating on the strand plasma (both radiative and thermal-conductive losses) and subsequent plasma draining versus the re-energisation arising from another nanoflare event. The fundamental cooling time of a strand (and hence the overall amalgamated loop) is directly proportional to the loop length \citep{barnes19}, which in this study remains fixed. Hence, similar time lags are observed for loops whose nanoflare repetition times are longer than this cooling time

Therefore, given the above behaviour, it is likely very difficult to conclusively infer the heating rate of sub-element strands from any single specific time-lag observation unless the heating repetition time is much shorter than the overall characteristic cooling timescales. However, there is an alternative approach that requires utilising the possible detection of several similar observed time lags across the EUV channel pairs but over an extended time period, at least say several times the estimated cooling timescale.

\begin{figure*}[!ht]
\centerline{\includegraphics[width=0.9\textwidth]{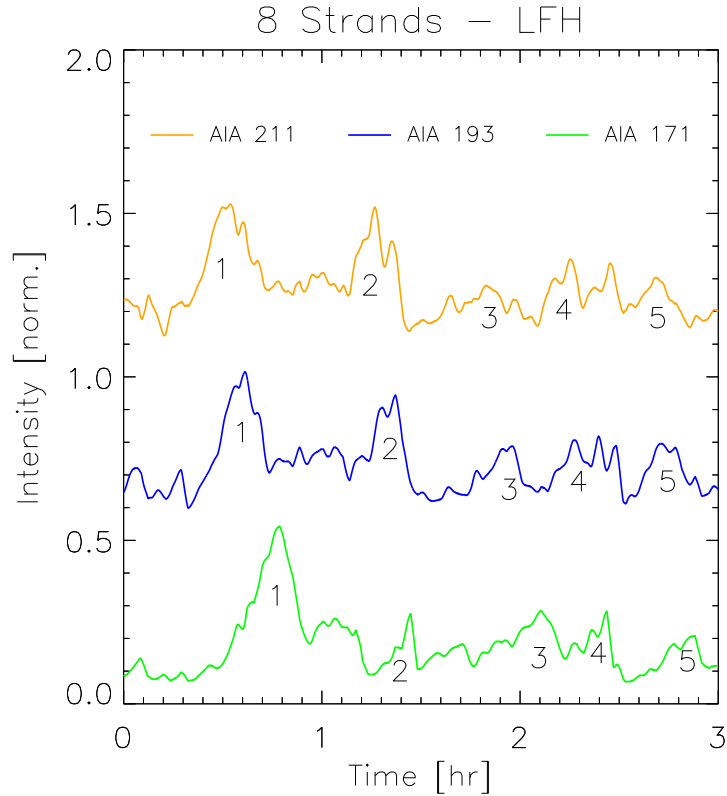}}
\caption{Synthetic AIA channel time profiles over an example three-hour period for the lowest (\textit{left}: 1500-second nanoflare repetition time on 8 strands) frequency heated loop scenario in this study for AIA 211 (\textit{orange}), 193 (\textit{blue}), and  171 (\textit{green}). The emission has been time-averaged for 120\,seconds and the five distinct rise-and-fall in emission from nanoflare heating are indexed in all three AIA channels.}
\label{fig:response}
\end{figure*}

Consider a simple, monolithic loop structure with a nanoflare repetition time greater than the loop cooling time; each nanoflare energy-release event in the loop would result in a rise and fall in the emission lines with the resulting calculated time lag of one event being close to the cooling time. However, if these time lags are observed repeatedly throughout the observational window, then the time between say, the onset of the observed time lag in each line from one determined time-lag period to the next would provide an estimation of the nanoflare repetition time.  However, the situation can become more involved when a multi-stranded approach is introduced, with the possibility of several nanoflares taking place nearly simultaneously and contributing to a confusing cumulative loop-emission profile. 

With this in mind, consider the three-hour observational window shown in Figure\,\ref{fig:response} for the least-frequently heated loop in the case study (eight strands subjected to LFH). Here, the light curves of AIA channels 211 (orange), 193 (blue) and 171 (green) are plotted and there are five distinct periods (numbered 1\,--\,5) of rise and fall in emission that are associated with a positive time lag that are observable in all three channels. Measuring the time lag between emission channel pairs from peak-to-peak yields a mean time lag of $792 \pm 146$\,seconds with a 95\,\% confidence level. As is seen from Figure\,\ref{fig:timelags}, this mean time lag corresponds well to the fundamental cooling timescale of the loop. Subsequently, calculating the repetition time between each of these time lags yields a value of $\approx1500$\,seconds, which is equivalent to the mean period of nanoflare repetition. Thus, measuring the period between a series of successive time lags that are greater than the fundamental cooling timescale of a particular loop may then be used as a tool to indicate how frequently the loop is being heated. 

However, with that said it is worth noting that additional peaks can be seen, such as the peak between 4 and 5 in AIA 211 in Figure\,\ref{fig:response}, which does not appear to correspond to any peaks in 193 and/or 171. This could be because this is an additional heating event that prolongs the duration for which the loop plasma remains at temperatures of $\approx$1.5\,MK and, as such, a single peak is seen rather than two. Additionally, the time-series plots shown in Figure\,\ref{fig:response} only analyse a segment of the loop at the apex. Hence, the plasma at a particular point within the loop subjected to a heating event seen in one channel could be swept to another portion of the loop by mass flow, and subsequently may not appear in the neighbouring AIA channels. For this reason, the visual inspection of Figure\,\ref{fig:response} only considers the peaks in intensity that can be traced through the three AIA channels shown. This highlights the complexity of analysing the light curves of individual events within multi-stranded loops, and thus any analysis doing so must proceed with caution.

\begin{figure*}[!ht]
\centerline{\includegraphics[width=1.25\textwidth]{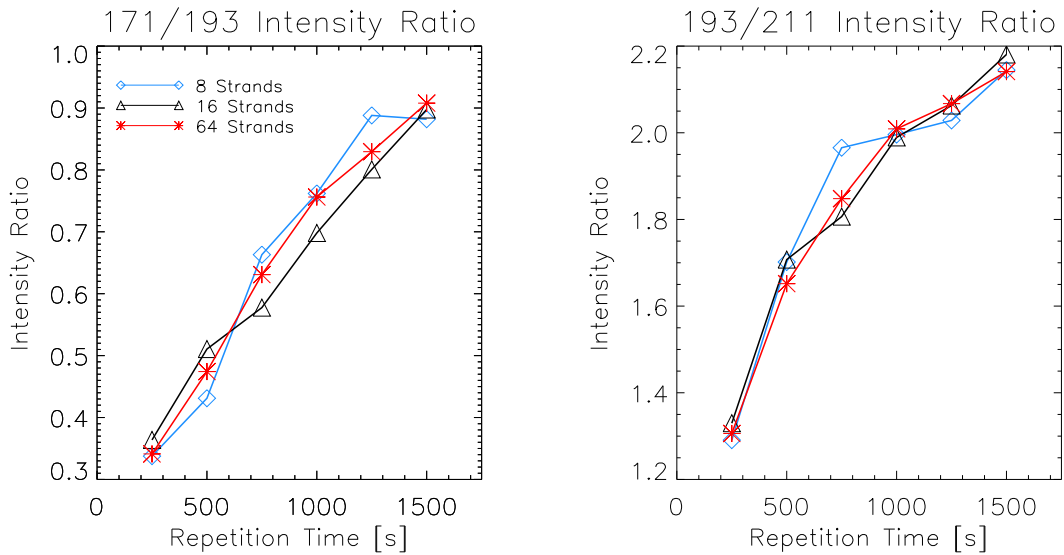}}
\caption{Emission-line ratios for AIA 171/193 (\textit{left}) and AIA 193/211 (\textit{right}) for 8- (\textit{blue}), 16- (\textit{black}), and 64-stranded loops (\textit{red}) shown as a function of nanoflare repetition time.}
\label{fig:intensity}
\end{figure*}
\subsection{AIA Emission Channel Ratios}
The mean AIA emission-channel ratios over the six-hour period analysed are shown in Figure\,\ref{fig:intensity} for 171/193 and 193/211 for all 18 loop configurations in the case study. Here, strong correlations are seen between the heating frequencies of each loop and their emission-channel ratios. As one would expect, strands that are subjected to fewer, but larger, magnitude nanoflares result in a loop that has greater 171/193 and 193/211 emission-channel ratios compared to a loop whose strands are subjected more frequently to lower magnitude nanoflares. This is because a loop that is heated more frequently has less opportunity to cool to lower temperatures before being re-energised compared to a less-frequently heated loop, as is evidenced by the standard deviation in apex temperature (Figure\,\ref{fig:std_dev}).

\subsubsection{Determining Heating Frequency and Strand Topology} 
In this subsection, a new technique is described that could prove a useful method for determining within a real, observed loop an estimate of the possible number of strands contained in the structure along with the subsequent heating frequency. This method combines AIA emission channel ratio (Figure\,\ref{fig:intensity}), standard deviation in loop apex temperature over a period of time (Figure\,\ref{fig:std_dev}), and a numerical model such as the multi-stranded loop code.

Forward modelling a loop with AIA observations alone, it is only possible to measure the emission channel ratios and thus it is only possible to constrain the nanoflare repetition time, as the number of strands appears to have little to no effect on the emission channel ratios (Figure\,\ref{fig:intensity}) for this study. In order to additionally constrain the number of strands that may be contained within a real AIA loop, co-aligned spectrometer data are needed such that a mean temperature and an associated standard deviation can be calculated and compared to the numerical models.

For example, consider a supposed observed loop that has a standard deviation in temperature of $\approx 0.4$\,MK over a six-hour time period with AIA 171/193 and 193/211 emission channel ratios of 0.92 and 2.12, respectively. From Figure\,\ref{fig:std_dev} there are three potential candidates that replicate the standard deviation in apex temperature, all of which have different heating frequencies and strand numbers (8 strands HFH, 16 strands ILFH, and 16 strands LFH). However, their emission channel ratios for 171/193 and 193/211 are all distinctly different (see Table\,\ref{tab:ratios}), and thus the model whose emission channel ratios most closely match the observed loop could potentially reveal the heating frequency and number of strands contained within a real loop. In this example, the real loop would likely be comprised of 16 sub-element strands that are subjected to LFH.
\begin{table}[!ht]
\caption{Emission channel ratios for forward modelling.}
\label{tab:ratios}
\begin{tabular}{c c c c c}
\hline 
No. of & Repetition & 171/193 & 193/211 & $\sigma_T$ \\    
Strands & Time [s] & Ratio & Ratio & [MK] \\ 
 \hline
8 & 500 & 0.43 & 1.71 & 0.38 \\
16 & 1250 & 0.80 & 2.07 & 0.39 \\
16 & 1500 & 0.90 & 2.18 & 0.41 \\
 \hline
\end{tabular}
\end{table}

In order for these comparisons between observations and simulations to be made, there are several factors that must first be considered. Firstly, estimates of the loop length and radius must be undertaken. The former could be determined by performing non-linear force free-field extrapolation of EUV imager data yielding the magnetic-field line geometry; the length of the magnetic-field lines can then be determined and used as a proxy for loop length. The latter can be determined by measuring the cross-sectional widths of emission profiles, such as is demonstrated by \citet{williams20a,williams20b}.

Using co-aligned spectrometer data, estimates of the loop density and temperature can be made over a period of time. From this, the mean loop-apex temperature and the standard deviation could be calculated, and when combined with the loop geometry this can be used to generate several models that replicate these properties with different numbers of sub-element strands and heating frequencies. Then the AIA-channel ratios can be determined for the real and forward-modelled loops, and as previously outlined can be compared to each other to estimate the number of strands and their most likely nanoflare repetition time(s). Finally, the inclusion of spectroscopic data would provide a further constraint on the forward-modelled loops by checking that the average (Doppler) velocities and densities of the observed loop are consistent with those generated by the model over the period analysed.

\section{Summary and Concluding Remarks}\label{sec:conc}
The multi-stranded loop code employed by \citet{sarkarwalsh08,sarkarwalsh09} has been updated to model the numerically challenging transition region as a discontinuity \citep{johnstonhood17a,johnstonhood17b}. This allows the multi-stranded loop code to better resolve the interaction of the transition region with down-flowing plasma as a result of a nanoflare event. It is demonstrated that the new code leads to an increase in coronal velocity and density compared to previous versions \citep{sarkarwalsh08,sarkarwalsh09} -- something that may often be underestimated in coronal-loop models in response to impulsive heating \citep{bradshawcargill13}.

With an under-resolved transition region, artificially low coronal densities can often be obtained because the downward heat flux ``jumps'' across the unresolved region to the chromosphere, underestimating the upflows (see, e.g., \citealp{bradshawcargill13,johnstonbradshaw19,johnston20}). However, it is demonstrated that the updated code leads to an increased evaporative upflow velocity when compared with the previous method. This results in a higher coronal density that shows good agreement with a fully resolved model, when subjected to impulsive heating \citep{bradshawcargill13}. The parameter space is explored for loops comprised of a low (8), intermediate (16), and high (64) number of strands that are subjected to a range of nanoflare repetition times, which are consistent with those specified by \citet{cargill15}.

Synthetic AIA loop-top data are generated and the light-curve time series are explored using the time lag analysis method \citep{viall11} for the cooler AIA channels (211, 193, and 171)\footnote{Wavelet analysis is also performed on the light curves, however the results proved inconclusive and as such have not been included in this article.}. As noted by \citet{viall16}, the time-lag technique will identify any cooling plasma present within a loop following an impulsive event by cross-correlating the rise and fall in emission of light-curve pairs. The light curves of an impulsively heated loop can result in a cyclical rise and fall of emission in light curves that might be detected and quantifiable using the time-lag method, although it has previously been demonstrated that the majority of time lags are consistent with cooling \citep{viall13}. The results presented in Figure\,\ref{fig:timelags} reveal that the time lag typically increases as the nanoflare repetition time (and the magnitude of energy release per impulsive event) increases. However, loops that consist of fewer sub-element strands display variation as large as 100\,--\,200\,seconds from one nanoflare repetition time to the next (compared to the 64-stranded loops). This variation makes it difficult to relate a particular time lag value to a given repetition time. For example, the loop consisting of eight strands that is subjected to a nanoflare every 1500\,seconds has a time lag of $\approx700$\,seconds, which makes it not possible to differentiate from several other scenarios in the case study (16 strands subjected to ILFH; 16 and 32 strands subjected to IHFH). Similarly, the gradient of the time lag vs. repetition-time plots flatten for repetition times $\gtrsim1000$\,seconds, and as such accurately determining the repetition time associated with a given time lag for even the 64-stranded loops is not viable.

With that said, an alternative approach to quantifying the nanoflare repetition times is outlined when the observed time lags greater than the fundamental cooling timescales of the loop in question. It is demonstrated in this article that measuring the duration between successive time lags in this scenario provides an approximate estimation of the input-heating frequency. However, it is worth noting that for this method to provide a reliable estimate it is required that the input-heating frequency is sufficiently low such that the plasma in the strands making up the loop is able to cool through the channels being analysed.

Comparing the AIA channel ratios of 171/193 and 193/211 for the 18 loops (Figure\,\ref{fig:intensity}) provides some promising initial results on obtaining the heating frequency of coronal loops. A strong correlation is found between the ratio values and nanoflare repetition times regardless of the multi-strandedness of the loop(s) in consideration. Comparing the AIA channel ratios with observations may prove to be a method by which the strand heating frequency can be inferred by forward modelling. Furthermore, combining the model loops and AIA observations with spectrometer data will allow for mean and standard deviation measurements of loop temperature and as shown in Table\,\ref{tab:ratios}, this standard-deviation and the AIA channel ratios can constrain the number of strands and their heating frequency.

For example, if an observed loop has a standard deviation in temperature of $\approx 0.4$\,MK, then from Figure\,\ref{fig:std_dev} there are three potential candidates that replicate this. However, their AIA channel ratios for 171/193 and 193/211 are all distinctly different (see Table\,\ref{tab:ratios}), and thus the model whose AIA channel ratios most closely match the observed loop could potentially reveal the heating frequency and number of strands contained within a real loop. Without co-aligned spectrometer data to complement AIA, it would not be possible to accurately determine the mean loop temperature and the associated standard deviation and thus an estimate of the number of sub-element strands could not occur. Additionally, given the large number of parameters at play (loop aspect ratio, loop length, temperature, geometry, Doppler/flow velocities, etc.) it is likely this analysis would need to be performed on a loop-by-loop basis. This will be explored further in a future study analysing \textit{Hinode}/EIS loops (see \citealp{xie17}).

As is evidenced in this article and discussed in detail by \citet{odwyer10}, there is currently poor spectroscopic coverage of plasma at temperatures $\gtrsim 2$\,MK. The upcoming NASA sounding rocket mission \textit{Marshall Grazing Incidence X-ray Spectrometer} (MaGIXS: \citealp{magixs,vigil21}) aims to fill this void by resolving soft X-ray spectra (Fe {\sc xvii}\,--\,Fe {\sc xx}) above a 4\,MK active region. Future work utilising the data from MaGIXS, as well as the X-ray and EUV spectrometer data from ESA's \textit{Solar Orbiter} is planned to further examine whether it is possible to determine the number of strands and/or their heating frequencies using the techniques presented in this article.

\begin{acks}
The Multi-Stranded Loops code is available for download on GitHub at \url{github.com/DrTomWilliams/Multi-Stranded-Loops-Code}. CDJ: The research leading to these results has received funding from the UK Science and Technology Facilities Council (consolidated Grants ST/N000609/1 and ST/S000402/1) and the European Union Horizon 2020 research and innovation programme (Grant agreement No. 647214).
\end{acks}

{\footnotesize\paragraph*{Disclosure of Potential Conflicts of Interest} The authors declare that they have no conflicts of interest.}

\bibliographystyle{spr-mp-sola}

\end{article} 
\end{document}